\documentclass[twocolumn]{aastex631}

\usepackage{comment}

\shorttitle{Merger of a Neutron Star with a Black Hole: one family vs. two-families scenario}
\usepackage{amsmath,amsfonts,amssymb,graphicx,color,times,bbm,psfrag,dsfont,slashed,bigints,bm}

\begin{document}

\title{Merger of a Neutron Star with a Black Hole: one-family vs. two-families scenario}

\author[0000-0002-8257-3819]{Francesco Di Clemente}
\affiliation{Dipartimento di Fisica e Scienze della Terra dell’Universit\`a di Ferrara, Via Saragat 1,
I-44122 Ferrara, Italy}
\affiliation{INFN Sezione di Ferrara,
Via Saragat 1, I-44122 Ferrara, Italy}

\author[0000-0003-1302-8566]{Alessandro Drago}
\affiliation{Dipartimento di Fisica e Scienze della Terra dell’Universit\`a di Ferrara, Via Saragat 1,
I-44122 Ferrara, Italy}
\affiliation{INFN Sezione di Ferrara,
Via Saragat 1, I-44122 Ferrara, Italy}

\author[0000-0003-3250-1398]{Giuseppe Pagliara}
\affiliation{Dipartimento di Fisica e Scienze della Terra dell’Universit\`a di Ferrara, Via Saragat 1,
I-44122 Ferrara, Italy}
\affiliation{INFN Sezione di Ferrara,
Via Saragat 1, I-44122 Ferrara, Italy}

\begin{abstract}
A kilonova signal is generally expected after a Black Hole - Neutron Star merger. The strength of the signal is related to the equation of state of neutron star matter and it increases with the stiffness of the latter. The recent results obtained by NICER from the analyses of PSR J0740+6620 suggest a rather stiff equation of state and the expected kilonova signal is therefore strong, at least if the mass of the Black Hole does not exceed $\sim 10 \mathrm{\,M}_\odot$ and if the adimensional spin parameter is not too small and the orbit is prograde. We compare the predictions obtained by considering equations of state of neutron star matter satisfying the most recent observations and assuming that only one family of compact stars exists with the results predicted in the two-families scenario. In the latter a soft hadronic equation of state produces very compact stellar objects while a rather stiff quark matter equation of state produces massive strange quark stars, satisfying NICER results. The expected kilonova signal in the two-families scenario is very weak: in particular the Hadronic Star - Black Hole merger produces a much weaker signal than in the one-family scenario because the hadronic equation of state is very soft. Moreover, according to the only existing simulation, the Strange Quark Star - Black Hole merger does not produce a kilonova signal because the amount of mass ejected is negligible.
These predictions will be easily tested with the new generation of detectors if Black Holes with an adimensional spin parameter $\chi_\mathrm{BH}\gtrsim 0.2$ or a mass $\mathrm M_\mathrm{BH} \lesssim 4 \mathrm M_\odot$ can be present in the merger.

\end{abstract}

\section{Introduction} \label{sec:intro}

Black Hole - Neutron Star (BH-NS) mergers are astrophysical phenomena of great interest because they not only produce gravitational wave (GW) signals, but they can also have very energetic electromagnetic (EM) counterparts in the form of short gamma ray bursts (sGRBs) and of kilonova (KN) explosions \citep{Shibata:2007zm}. The disruption of the NS produces the dynamical ejection of some material and the formation of a disk of hot matter around the BH and, in turn, these processes can be at the origin of a sGRB and a KN signal. In order to produce these EM signals the BH should not be too massive otherwise it is not possible to form an accretion disk and to eject material. Instead, when the BH mass is $\lesssim 10 \, \mathrm{M_\odot}$, the adimensional spin parameter is sufficiently high and the equation of state (EoS) of the compact star is not too soft, a mass up to a few $0.01 \mathrm{\,M}_\odot$ is dynamically ejected and a larger mass, up to a few $0.1 \mathrm{\,M}_\odot$, forms an accretion disk which can later be ablated by the neutrinos allowing a further ejection of mass. 

It is very important to note that the latest results from the NICER analyses of PSR J0740+6620 indicate that the EoS of NS matter is rather stiff, at least for the most massive compact stars \citep{riley2021nicer,miller2021nicer,Raaijmakers:2021uju}. If only one family of compact stars exists we can therefore expect a strong KN signal associated with many BH-NS mergers. 

In the two-families scenario the phenomenology of BH-NS mergers can be rather different.
Within that scenario, outlined in \citet{drago_can_2014}, hadronic stars (HSs) and strange quark stars (QSs) coexist; the EoS of hadronic matter is rather soft due to the formation of hyperons and delta resonances at densities larger than about twice nuclear saturation density. In turn, this allows for the existence of HSs having small radii and a mass not exceeding $\sim 1.6$ M$_{\odot}$. The EoS of quark matter, instead, can be rather stiff and the QSs branch is populated by large and massive objects, having masses that can potentially reach $2.6$ M$_{\odot}$, see \citet{bombaci_was_2021}. 

In the two-families scenario a key assumption is the validity of the Bodmer-Witten hypothesis on the absolute stability of strange quark matter. Namely, at zero pressure strange quark matter is more bound than iron. In turn, this implies that hadronic matter is metastable and it "decays", under certain conditions, into strange quark matter. 
The conditions for such a conversion are related to the amount of strangeness which is present in hadronic matter \citep{Bombaci_2004,drago_can_2014}. In \citet{pietri_merger_2019} it has been estimated that only when the hyperon fraction exceeds $\sim 0.1$ can the quark phase start to be produced via nucleation. This occurs at densities of a few times the saturation density. 
Thus, HSs with a central value below this threshold are actually stable. Only HSs with larger values of the central density can convert into strange QSs. 
In this scenario therefore HSs and QSs coexist and populate two different branches of compact stars.
The possibility of forming a QS with a radius larger than that of the HS having the same baryonic mass is a special feature of the two-families scenario and the underlying dynamics have been clarified in many papers and in particular in \citet{Drago:2020gqn}. 

In the two-families scenario the outcome of the merger of a BH and a compact star clearly depends on the nature of the low mass companion.
If the low mass companion is a QS, the numerical simulations of \citet{Kluzniak:2002dm} suggest that no significant amount of material is dynamically ejected or left in the accretion torus. 

The two-families scenario has been developed in order to account for the possible existence of very compact stars, having a radius $\lesssim 11.5 \mathrm{\,km}$ for a mass of $\sim (1.4-1.5)\, \mathrm{M_\odot}$. It has been shown (see e.g. \citet{Most:2018hfd}) that it is not possible to obtain such small radii in the absence of a strong phase transition, as that present in the two-families scenario. This scenario therefore becomes phenomenologically irrelevant if all compact stars have radii $\gtrsim 11.5 \mathrm{\,km}$. Since the direct measurement of radii is non trivial and is affected by large systematic errors, it is important to find alternative ways to test the existence of stars with very small radii. This is the aim of this paper, in which we show that stars having very small radii produce a significantly suppressed KN signal.

In this paper we will compare the predictions for the KN signal generated in a BH-NS merger by assuming either that:
\begin{itemize}
    \item only one family of NSs exists and that it satisfies the most recent observational constraints,  or that 
    \item two-families of compact stars exist and the merger of the BH is with a HS (first family), since the QS-BH merger will be assumed not to generate a KN.
\end{itemize}

\section{Semi-analytical model}

In order to get an estimate of the mass ejected we use the semi-analytical models of \cite{Barbieri2020,Foucart2018,Kawaguchi2016} which provides a fit to the data obtained in the simulations of BH-NS mergers. The models allow us to estimate the mass of the disk $M_{\mathrm{disk}}$ and the dynamical ejecta mass $M_\mathrm{dyn}$ in terms of five quantities: the mass, compactness and tidal deformability of the NS ($M_{\mathrm{NS}}$, $C_\mathrm{NS}$ and $\Lambda_{\mathrm{NS}}$) and the mass and the parallel spin component of the BH ($M_{\mathrm{BH}}$ and $\chi_{\mathrm{BH,||}}$). Once $M_{\mathrm{disk}}$ and $M_\mathrm{dyn}$ are estimated it is possible to predict the strength of the KN signal \citep{Barbieri2020}.

The total mass of matter not immediately absorbed by the BH, $M_{\mathrm{out}}$, is the sum of two components: $M_{\mathrm{disk}}$, representing
the gravitationally bound material, and $M_{\mathrm{dyn}}$,
the unbound part. $M_{\mathrm{out}}$ is given by an interpolation formula as \citep{Foucart2018}
\begin{equation}
M_\mathrm{out} = M^\mathrm{b}_\mathrm{NS}\left[\mathrm{max}\left(\alpha\frac{1-2\rho}{\eta^{1/3}}-\beta\tilde{R}_\mathrm{ISCO}\frac{\rho}{\eta}+\gamma,0\right)\right]^\delta
\label{eq:mout}
\end{equation}
where $\alpha$, $\beta$, $\gamma$, $\delta$ are fitting parameters. In the analyses of \citet{Foucart2018} they make use of the so-called symmetric mass ratio and in this way the parametrization remains stable even for masses of the BH and of the NS which are comparable. Notice also that in our analyses we are using mass ratios well inside the range of validity of the parametrization obtained in \citet{Foucart2018}. Therefore there is no dependence of the value of the parameters on the NS mass.

In the formula above $M^\mathrm{b}_\mathrm{NS}$ is the NS baryonic mass and $\rho=(15\Lambda_\mathrm{NS})^{-1/5}$ is a function of the tidal deformability $\Lambda_\mathrm{NS}$. $\eta$ is the symmetric mass ratio defined as
\begin{equation}
\eta=q/(1+q)^2 \, ,
\label{eq:symm_mass_ratio}
\end{equation}
 where $q={M_{\mathrm {NS}}}/{M_{\mathrm {BH}}}$ is the mass ratio. $\tilde{R}_\mathrm{ISCO}=R_\mathrm{ISCO}\,c^2/GM_\mathrm{BH}$ is the dimensionless ISCO (Innermost Stable Circular Orbit). This quantity is defined in \cite{bardeen1972} as:
\begin{align}
\begin{aligned}
\tilde{R}_{\mathrm{ISCO}}(&\chi)  = 3+Z_2(\chi)+\\ &-{\mathrm{sgn}}(\chi)\sqrt{(3-Z_1(\chi))(3+Z_1(\chi)+2Z_2(\chi))}
\end{aligned}
\end{align}
where 
\begin{equation}
    Z_1(\chi)=1+(1-\chi^2)^{1/3}[(1+\chi)^{1/3}+(1-\chi)^{1/3}]
\end{equation}  
and 
\begin{equation}
    Z_2(\chi)=(3\chi^2+Z_1(\chi)^2)^{1/2} \, .
\end{equation}

In Equation \ref{eq:mout} parameters are fixed and don't depend on the NS mass, since the contribute of this mass in encoded in $\rho$ and $\eta$. Moreover. $R_\mathrm{ISCO}$ is the BH ISCO, since in the original derivation the behaviour of the unbound material is extrapolated assuming $M_{\mathrm{BH}}/M_{\mathrm{NS}} \rightarrow \infty$.

The dynamical ejecta mass is instead approximated by: 
\begin{align}
\begin{aligned}
M_\mathrm{dyn}&=M^\mathrm{b}_\mathrm{NS}\Big\lbrace\mathrm{max}\left[ a_1q^{-n_1}(1-2C_\mathrm{NS})/C_\mathrm{NS} \right.+\\&- a_2q^{-n_2}\tilde{R}_\mathrm{ISCO}(\chi_\mathrm{BH,||})+\\& + \left.a_3(1-M_\mathrm{NS}/M^\mathrm{b}_\mathrm{NS})+a_4,0 \right] \Big\rbrace\label{eq:ejected}
\end{aligned}
\end{align}
where $a_1$, $a_2$, $a_3$, $a_4$, $n_1$, $n_2$ are the fitting parameters. $\chi_\mathrm{BH,||}=\chi_\mathrm{BH}\cos{\iota_\mathrm{tilt}}$ is the BH parallel spin component which depends on the adimensional BH spin $\chi_\mathrm{BH}$ and on $\iota$, the angle between the BH spin and the total angular momentum. Notice that to have a large value for $M_{\mathrm{dyn}}$ the orbit must be prograde with respect to the BH spin. Here and in the following, we assume therefore prograde orbits i.e. $\chi_\mathrm{BH,||}\geq 0$.
It is then possible to estimate the mass of the accretion disk as:
\begin{equation}
M_\mathrm{disk}=\mathrm{max}\left[ M_{\mathrm{out}} - M_{\mathrm{dyn}},0\right] 
\end{equation}
i.e. the bound material is the total material outside the BH minus the gravitationally bound part.

Following \citet{Barbieri2020} we set the limit for the dynamical ejecta mass as 
\begin{equation}
    M_\mathrm{dyn,max}= f \, M_\mathrm{out} \, ,
    \label{eq:max_ejected}
\end{equation}
where $f$ is the maximum ratio between the dynamical ejecta mass and the total mass outside the BH.

In conclusion, after an EoS for NS matter has been selected, both $M_\mathrm{disk}$ and $M_{\mathrm{dyn}}$ can be evaluated as functions of
$M_{\mathrm {NS}}$, $M_{\mathrm {BH}}$ and  
$\chi_\mathrm{eff}$.

\section{Observational limits on the Equation of State}

\begin{figure}[h!]
\begin{minipage}{0.5\textwidth}
\includegraphics[width=\textwidth]{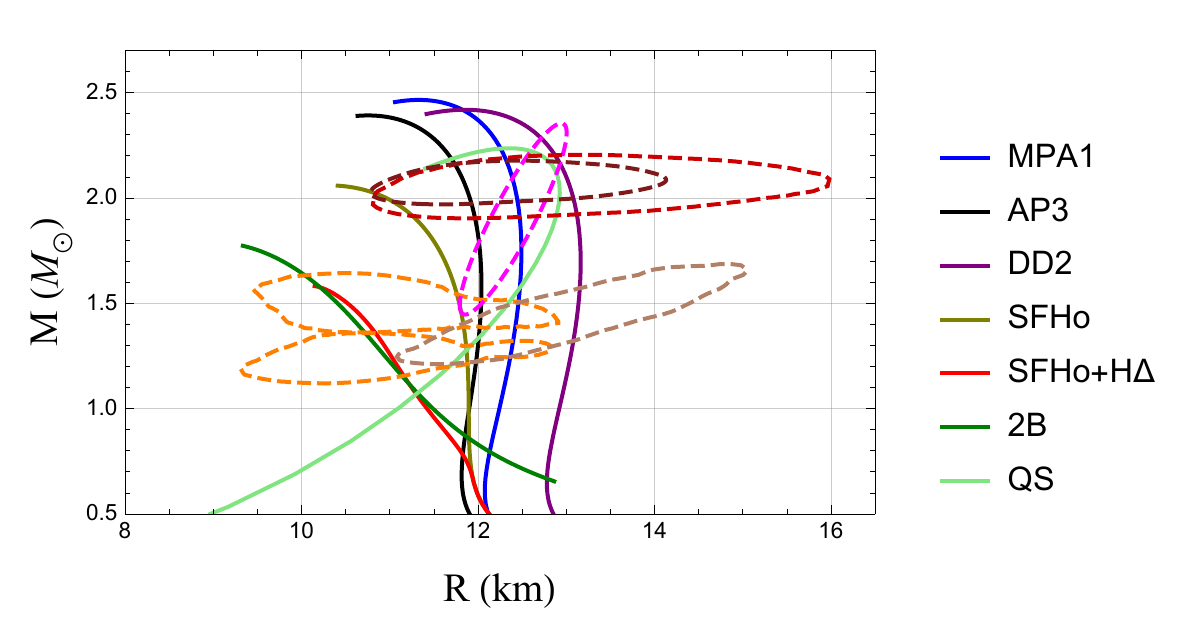}
\end{minipage}
\begin{minipage}{0.5\textwidth}
\includegraphics[width=0.95\textwidth]{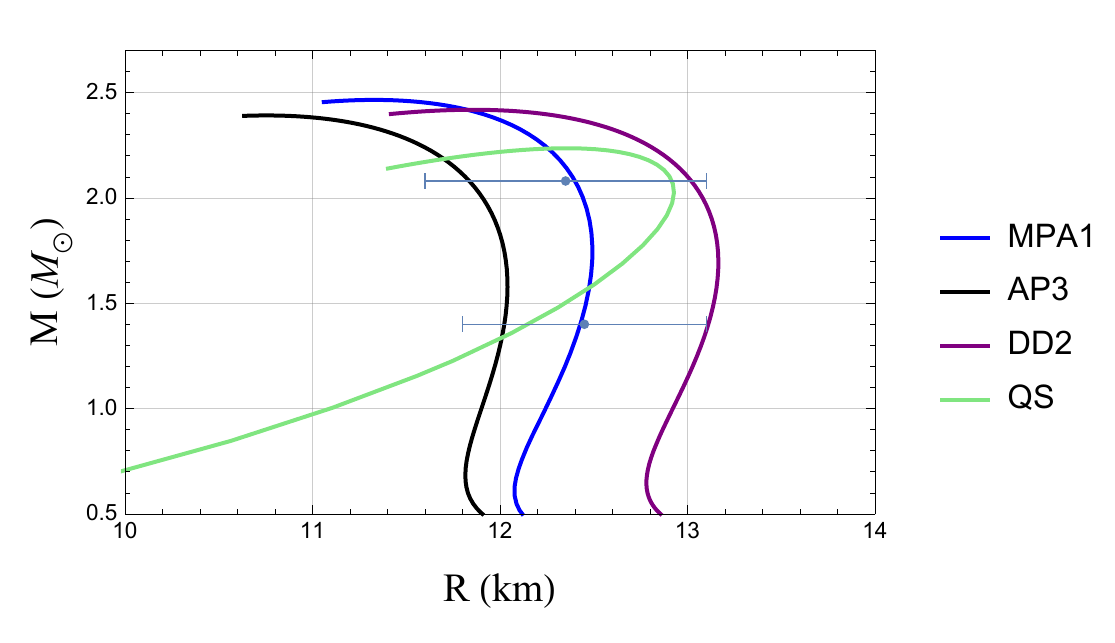}
\end{minipage}
\caption{Upper panel: observational limits on masses and radii of selected sources (dashed lines), compared with a few EoSs (solid lines). NICER results for PSR J0740+6620: brown from \citet{riley2021nicer} and dark red from \citet{miller2021nicer}. NICER results for PSR J0030+0451, sepia, from \citet{Riley:2019yda}. Violet, limits on 4U 1702-429 from \cite{Nattila:2017wtj}. Orange, limits from GW170817 from \citet{Abbott:2018exr}. Lower panel: limits on the radius at 68$\%$ of credibility interval for stars with masses $1.4 \mathrm{\,M_{\odot}}$ and $2.08 \mathrm{\,M_{\odot}}$ based on the analysis of NICER results and on GW170817 \citep{miller2021nicer}, with three nucleonic EoSs (used in our analysis) and a QS. The nucleonic EoSs are MPA1 \citep{MUTHER1987469}, DD2 \citep{typelDD2}, AP3 \citep{AP3}, SFHO \citep{sfho}. SFHO+HD \citep{PhysRevC.90.065809} incorporates $\Delta$-resonances and hyperons and 2B is a soft piece-wise polytropic used as a reference \citep{Markakis_2009}} 
\label{fig:mrconstraints}
\end{figure}

 The recent results of NICER indicate rather large radii for masses ranging from $\sim 1.4 \mathrm{\,M}_\odot$ up to $\sim 2 \mathrm{\,M}_\odot$. In the left panel of Fig.~ \ref{fig:mrconstraints} we show a few recent limits on masses and radii, and in particular those obtained by NICER. As it can be seen, EoSs moderately soft as SFHO are only marginally compatible with the data, which instead suggest either a stiff nucleonic EoS or a pure quark matter EoS (see \citet{Traversi:2021fad}). We also show two EoSs, 2B and SFHO+HD, which are NOT compatible with the limits presented in the figure. SFHO+HD is a hadronic EoS incorporating $\Delta$-resonances and hyperons and it is representative of the hadronic branch of the two-families scenario. The compact objects associated with this branch have small radii, as the ones suggested e.g. by \citet{Ozel:2016oaf,Capano:2019eae}. 2B is a simple piece-wise polytropic EoS \citep{Markakis_2009} and it has been used as a reference by \citet{Barbieri2020} to provide an example of soft EoS which does not produce a strong KN signal. It is important to notice that 2B is only slightly softer than SFHO+HD.  In the right panel of Fig. \ref{fig:mrconstraints} we compare the limits obtained by \citet{miller2021nicer} with the results of three purely nucleonic EoSs which are representative of the range of values of radii compatible with the observations, if only one family of compact stars exists. Two of the EoSs, MPA1 and DD2 have been discussed also in \citet{Barbieri2020}, while AP3 is close to the left limit indicated by \citet{miller2021nicer}. It is important to recall that, if only one family of compact stars exists, there is a rather precise linear relation between radius and tidal deformability of NSs having masses of about $1.5 \mathrm{\,M}_\odot$ \citep{Burgio:2018yix} and therefore the limits on the radii directly translate into limits on the tidal deformability.
 
\begin{figure}[h!]
\centering
\includegraphics[width=0.5\textwidth]{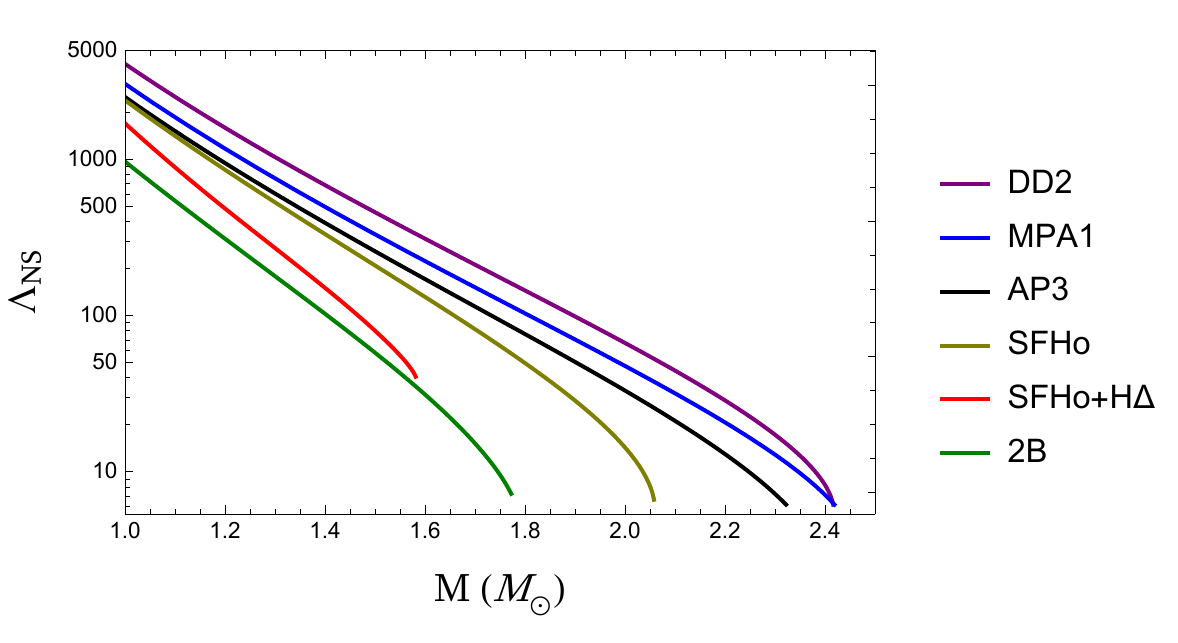}
\caption{Tidal deformability for various representative EoSs as a function of the NS mass. \label{fig:tidal}  }
\end{figure}

In Fig.\ref{fig:tidal} we show the tidal deformabilities for a similar set of EoS. Notice that the group of EoSs satisfying the limits of \citet{miller2021nicer} in the one-family case, are rather well separated from SFHO+HD and 2B, EoSs which can be justified in a two-family scenario.

\section{Predicted values of $M_\mathrm{disk}$ and of $M_{\mathrm{dyn}}$}

In previous papers \citep{Shibata:2007zm,Barbieri2020} it has been shown that a strong KN signal can be obtained if the EoS is stiff, due to large values for 
$M_\mathrm{disk}$ and $M_{\mathrm{dyn}}$.
Here we compare the estimated values of these two masses, computed by assuming that only one family of compact stars exists (and the EoS has therefore to satisfy limits of the type discussed in \citet{miller2021nicer}) with the values obtained in the two-families scenario for the merger of a HS with a BH.

\begin{figure*}[t] 
\begin{minipage}{0.7\textwidth}
\hspace{1.5 cm}\includegraphics[width=1\textwidth]{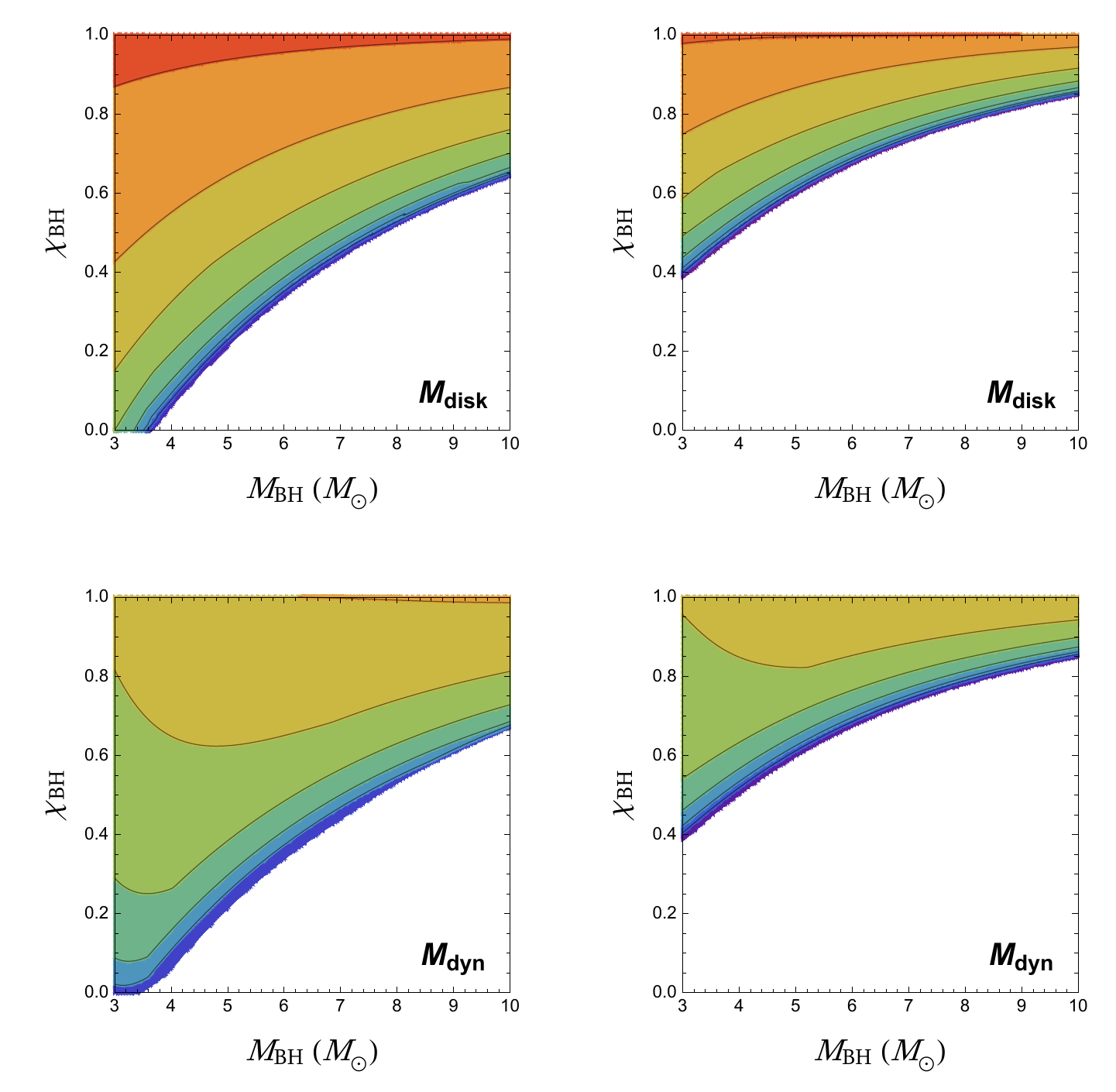}
\end{minipage}
\begin{minipage}{0.3\textwidth}
\hspace{2.3 cm}\includegraphics[width=0.35\textwidth,scale=0.3]{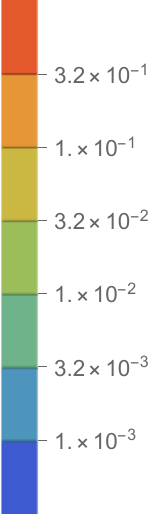}
\end{minipage}
 \caption{Plots for the mass of the disk on top and for dynamical ejecta on bottom. Left figures are relative to MPA1, right figures to SFHO+HD. The considered mass of the star is $\sim1.4 \mathrm{\,M_{\odot}}$. Values for tidal deformability for MPA1 and SFHO+HD are respectively  $\Lambda_{\mathrm{NS}} \simeq 462$ and $\Lambda_{\mathrm{NS}} \simeq 151$. Plots are function of the BH mass ($M_{\mathrm{BH}}$) and of the adimensional spin parameter $\chi_{\mathrm{BH}}$.}
 \label{fig:mpa1sfho}
  \end{figure*}

In Fig. \ref{fig:mpa1sfho} we compare the results obtained by using MPA1 with those obtained using SFHO+HD, for $M_{\mathrm{NS}}=1.4 \mathrm{\,M}_\odot$. Notice that the M-R relation based on MPA1 passes close to the central values obtained by the analysis of \citet{miller2021nicer}, as shown in the lower panel of Fig. ~\ref{fig:mrconstraints}. For simplicity, we have assumed $\iota_\mathrm{tilt}=0$. 
It is clear from Fig. \ref{fig:mpa1sfho} that when using MPA1 there exists a rather extended range of values of $\chi_{\mathrm{BH}}$ and of $M_{\mathrm{BH}}$ leading to large values of $M_\mathrm{disk}$ and $M_{\mathrm{dyn}}$, and therefore to strong KN signals, while for the same values of $\chi_{\mathrm{BH}}$ and of $M_{\mathrm{BH}}$ no mass is ejected if SFHO+HD is used.
The difference is particularly strong and relevant for small values of $M_{\mathrm{BH}}$; for instance if $M_{\mathrm{BH}}=4 \mathrm{\,M}_\odot$ no disk forms when using SFHO+HD for $\chi_{\mathrm{BH}}\lesssim 0.65$, while for MPA1 a disk forms for $\chi_{\mathrm{BH}}\gtrsim 0.3$. If one examines $M_{\mathrm{dyn}}$, the differences between the two scenarios are also present but less marked.

\begin{figure}[h!] 
    \centering
    \includegraphics[width=0.5\textwidth]{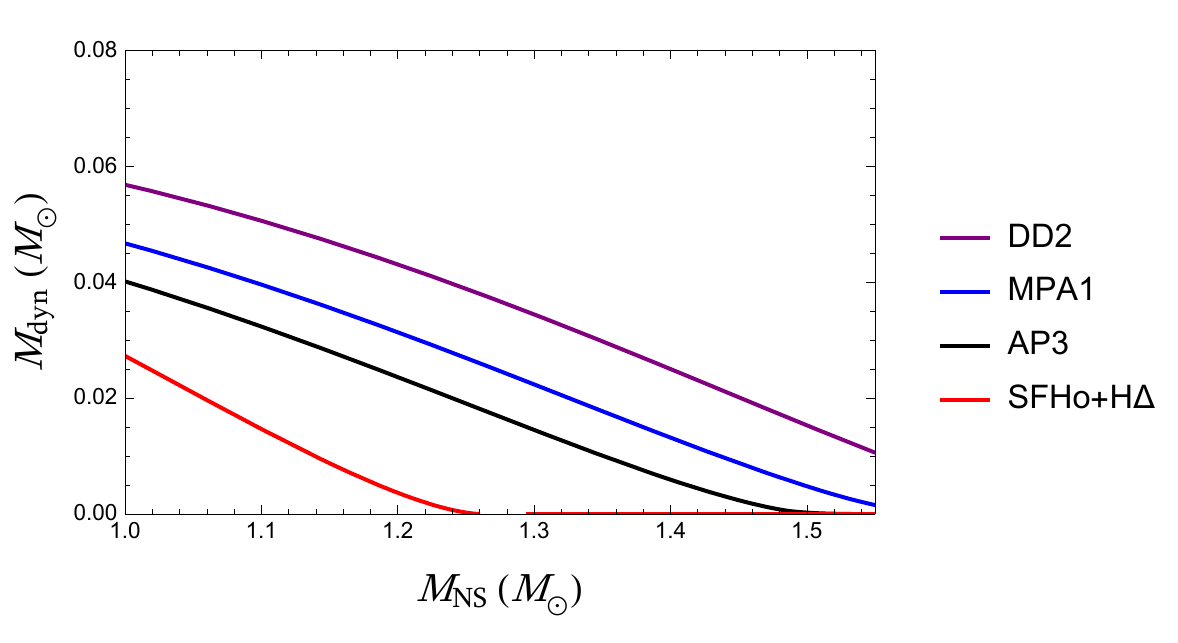}
  \label{fig:504}
  \caption{Dynamically ejecta mass for a BH of $5 \mathrm{\,M_{\odot}}$ with a spin parameter $\chi = 0.4$, as a function of the NS mass.  } \label{fig:mdyn_vs_mns}
\end{figure}

In Fig. \ref{fig:504} we compare the one-family vs the two-families scenario considering the three nucleonic EoSs presented in the lower panel of Fig.\ref{fig:mrconstraints}: their M-R relations are representative of the entire range of values allowed by the analysis of \citet{miller2021nicer}. For $M_{\mathrm NS}\sim (1.2-1.3) \mathrm{\,M}_\odot$ when using SFHO+HD the amount of mass dynamically ejected is much smaller than in the one-family scenario and it becomes 0 at $1.4 \mathrm{\,M}_\odot$. 

\section{Modelling observations}\label{sec:kn_model}

\subsection{A toy-model to mimic correlations between observables}\label{subsect:toymodel}
Gravitational-wave observational data from LIGO-Virgo (LV hereafter) provide a fairly accurate measurement of the chirp mass, but not an equally accurate measurement of the spin and individual masses of the components of a merger. When performing the data analysis, the values of the masses and spins turn out to be strongly correlated.  Therefore, we relied on a toy-model developed in \citet{ngetal}, that provides synthetic posteriors in order to emulate the data analysis.
The model shows how Gaussian and uncorrelated likelihoods for the symmetric mass ratio and the 1.5PN phase term (the quantity $\psi$ described below) can result in a skewed posterior for the effective spin parameter of the binary system. The resulting likelihood for the masses and the effective spin parameter reads

\begin{align}
\begin{aligned}
\mathcal{L}(M_{\mathrm{NS}},M_{\mathrm{BH}},\chi_{\mathrm{eff}})&=\mathcal{N}(\psi(M_{\mathrm{NS}},M_{\mathrm{BH}},\chi_{\mathrm{eff}});\psi_0,\sigma_{\psi}) \times \\& \mathcal{N}(\eta(M_{\mathrm{NS}},M_{\mathrm{BH}});\eta_0,\sigma_{\eta}) 
\label{eq:toy_model_1}
\end{aligned}
\end{align}
where the effective inspiral spin parameter is
\begin{equation}
\chi_\mathrm{eff}=\left(\frac{M_\mathrm{NS}}{M_\mathrm{NS}+M_\mathrm{BH}}\vec{\chi}_\mathrm{NS} + \frac{M_\mathrm{BH}}{M_\mathrm{NS}+M_\mathrm{BH}}\vec{\chi}_\mathrm{BH} \right) \cdot \hat{L} 
\end{equation}
in which $\hat{L}$ is the unit vector along the orbital angular momentum, $\vec{\chi}_\mathrm{NS}$ and $\vec{\chi}_\mathrm{BH}$ the NS and BH adimensional spin vectors.
In Equation (\ref{eq:toy_model_1}), $\mathcal{N}(x;x_0,\sigma_{x})$ represents a Gaussian in the variable $x$, centered in $x_0$ with standard deviation $\sigma_x$. The variable $\psi$ is defined as
\begin{equation}
\psi  =   \eta^{-3/5}\left[  \frac{(113 - 76 \eta) \chi_{\mathrm{eff}} + 76 \, \delta \, \eta \,  \chi_a}{128} - \frac{3\pi}{8}\right] \,.
\label{eq:psi}
\end{equation}
In the formula above $\delta = (M_\mathrm{BH} -M_\mathrm{NS})/(M_\mathrm{BH}+M_\mathrm{NS})$ and $\chi_a = (\chi_{\mathrm{BH,||}} - \chi_{\mathrm{NS,||}})/2$ where $\chi_{\mathrm{BH,||}}$ and $\chi_{\mathrm{NS,||}}$ are the parallel component of the spins. The spin of the less massive body is neglected by setting $\chi_{\mathrm{NS,||}} =0$, so that
\begin{equation}
\chi_{\mathrm{eff}} = \chi_{\mathrm{BH,||}}/(1+q)
\end{equation}
and
\begin{equation}
\chi_a = (1+q)\chi_{\mathrm{eff}}/2 \, .
\end{equation}

Finally, in our analysis we have taken into account the observational constraint on the chirp mass by adding a multiplicative term to Equation (\ref{eq:toy_model_1}):

\begin{align}
\begin{aligned}
\mathcal{L_{\mathrm{total}}}(M_{\mathrm{NS}},&M_{\mathrm{BH}},\chi_{\mathrm{eff}})=\mathcal{L}(M_{\mathrm{NS}},M_{\mathrm{BH}},\chi_{\mathrm{eff}}) \times \\& \mathcal{N}(M_{\mathrm{chirp}}(M_{\mathrm{NS}},M_{\mathrm{BH}});M_{\mathrm{chirp},0},\sigma_{M_{\mathrm{chirp}}}) \, .
\label{eq:toy_model}
\end{aligned}
\end{align}

\begin{figure}[h!]
\begin{minipage}{0.5\textwidth}
\includegraphics[width=0.8\textwidth]{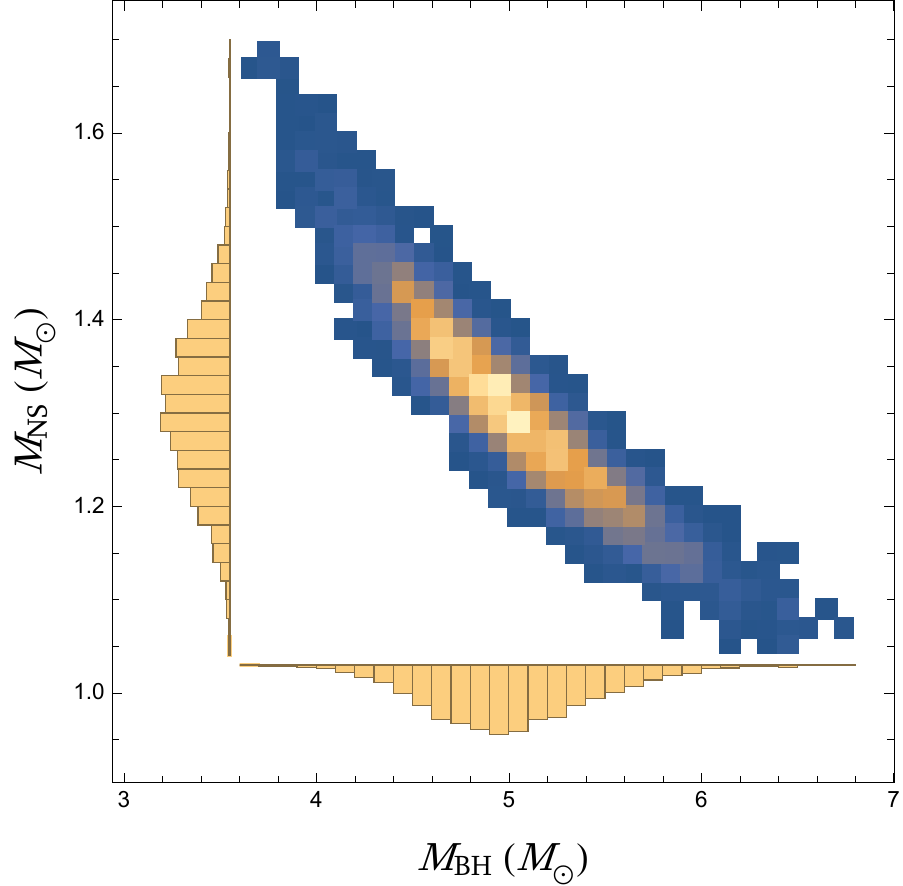}
\end{minipage}
\begin{minipage}{0.5\textwidth}
\includegraphics[width=0.8\textwidth]{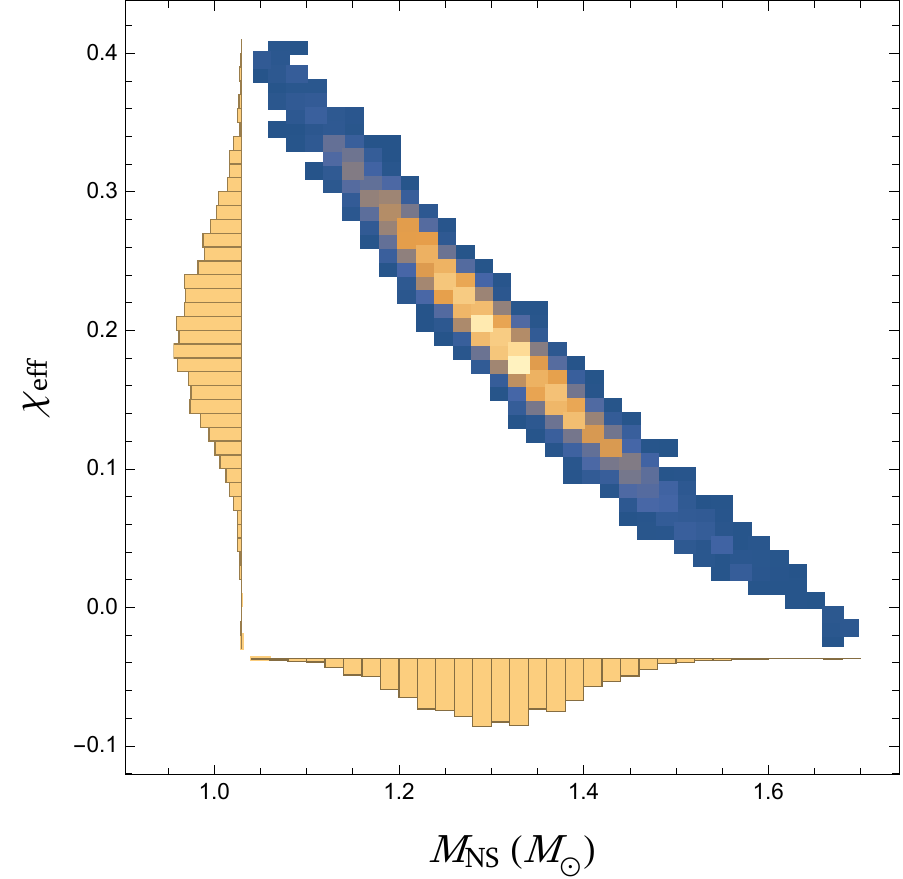}
\end{minipage}
\caption{Example of correlations obtained with the toy-model for an event characterized by the following values: $M_\mathrm{NS,0}=1.3\,\mathrm{M_\odot}$, $M_\mathrm{BH,0}=5\,\mathrm{M_\odot}$,  $\chi_{\mathrm{eff,0}}=0.2$, $\sigma_{\psi}=0.005$, $\sigma_{\eta}=0.015$ and $\sigma_{M_{\mathrm{chirp}}}=0.025$. Top: correlation between $M_\mathrm{NS}$ and $M_\mathrm{BH}$. Bottom: correlation between $M_\mathrm{NS}$ and $\chi_{\mathrm{eff}}$.} 
\label{fig:correlation}
\end{figure}

In our analyses, we identify an event by choosing the central values $M_\mathrm{NS,0}$, $M_\mathrm{BH,0}$ and $\chi_\mathrm{eff,0}$ and in this way we fix the values of $\psi_0$, $\eta_0$ and $M_\mathrm{chirp,0}$. The standard deviations $\sigma_{\psi}$, $\sigma_{\eta}$ and $\sigma_{M_{\mathrm{chirp}}}$ have been fixed in order to approximate the correlations observed in LV analyses and displayed in Fig. 4 and 8 of \citet{LIGOScientific:2021qlt}. In our Fig. \ref{fig:correlation}, the marginalized distribution functions obtained with a Markov chain Monte Carlo sampling of the likelihood of Equation (\ref{eq:toy_model}) are shown.

\subsection{Probability of observing the Kilonova signal}

To calculate the luminosity of the possible KN signal we rely on the model developed by \citet{Kawaguchi2016}. 
Since we know the limiting magnitude per band of LSST (Vera Rubin Observatory) \citep{chase_kilonova_2021}, we can calculate for each event and for each EoS
the probability of observing a KN signal. 
We set the distance of the hypothetical events at 200 Mpc, which is within the observing range of LV. 

The procedure we follow is: 
for each event we generate an ensemble of points according to Equation (\ref{eq:toy_model}); for each generated point we compute $M_\mathrm{dyn}$ by using Equations (\ref{eq:ejected}) and (\ref{eq:max_ejected}); we then calculate the bolometric luminosities, the bolometric magnitudes and the bolometric corrections\footnote{The bolometric correction of a specific observational band is the difference between the bolometric magnitude and the visible magnitude in that band.} for a single band filter (g-band filter); finally we compute the fraction of the sample which generates a visible magnitude smaller than the limiting one of LSST and in this way we obtain the probabilities displayed in Table \ref{tab:confidence}. 

\section{Results and Conclusions}\label{sec:kn_results}

As shown in Table \ref{tab:confidence}, the probability of observing a KN signal is negligible if the BH spin $\chi_\mathrm{BH}$ is close to 0. 
For instance, the event \textit{13ns7bh0c\textunderscore1s} is similar to GW200115 and we can confirm the result of other authors \citep{Zhu_2021,LIGOScientific:2021qlt} of a very low probability of observing a KN signal for that event even assuming a stiff EoS. Notice anyway that if the BH mass is very small ($\chi_\mathrm{BH} \lesssim 5 \mathrm M_\odot)$ and the EoS of the NS is particularly stiff a KN signal is expected even for not rotating BHs.

While our paper was in preparation, two other works appeared discussing the probability of observing a KN signal in a NS-BH merger, \citet{Zhu_2021} and \citet{Fragione_2021}.
It is important to clarify the differences between the approach followed in our work respect to those papers. First, \citet{Zhu_2021} discuss events generated by using the population synthesis code STARTRACK of \citet{Belczynski_2008,Belczynski_2020}. In that way they conclude that $\chi_{\mathrm{BH}}$ is smaller than about 0.2 and therefore the possibility of generating an observable KN signal is marginal. \citet{Fragione_2021} also makes use of various population synthesis results in order to estimate the probability of having mergers with given values of masses and spins and from those values it estimates the probability of generating an observable KN signal. In our approach we do not use population synthesis, but we concentrate on the information that can be obtained from the realistic analysis of an event in which the extrapolated values of the masses and of the spin are correlated. The reason we do not make use of population synthesis is that in the existing codes the possibility of having a NS-BH merger from a hierarchical triple system is not included \citep{Wiktorowicz}. On the other hand, the possibility of a merger originating from a triple system has been suggested e.g. in connection with GW190814 \citep{10.1093/mnras/staa3372,10.1093/mnras/stab178}. If the BH participating to the NS-BH merger is obtained from a previous merger of two NSs its mass can be small and it can be rapidly rotating.
Therefore, in the analysis presented in Table \ref{tab:confidence} we also discuss BHs which are more rapidly rotating.

\begin {table}[]
\centering
\begin {tabular} {||c|cccc||}
\hline
& SFHO+HD& AP3& MPA1& DD2 \\
\hline\hline
13ns5bh0c\_1s& 0.01& 0.13& 0.26& 0.48\\
\hline
13ns5bh0c\_05s& 0.00& 0.04& 0.18& 0.52\\
\hline\hline
13ns7bh0c\_1s& 0.00& 0.00& 0.00& 0.05\\
\hline
13ns7bh0c\_05s& 0.00& 0.00& 0.00& 0.00\\
\hline\hline
13ns5bh2c\_1s& 0.10& 0.53& 0.67& 0.83\\
\hline
13ns5bh2c\_05s& 0.02& 0.55& 0.79& 0.96\\
\hline\hline
13ns7bh2c\_1s& 0.00& 0.08& 0.19& 0.36\\
\hline
13ns7bh2c\_05s& 0.00& 0.02& 0.07& 0.36\\
\hline\hline
13ns5bh5c\_1s& 0.64& 0.95& 0.97& 0.99\\
\hline
13ns5bh5c\_05s& 0.82& 1.00& 1.00& 1.00\\
\hline\hline
13ns7bh5c\_1s& 0.23& 0.63& 0.72& 0.81\\
\hline 
13ns7bh5c\_05s& 0.15& 0.84& 0.97& 1.00\\
\hline\hline
\end {tabular}
\caption{Probability of observing a KN signal in the g-band by LSST after 1 day from the merger event, for four EoSs at a distance of 200 Mpc. The g-band limiting magnitude (AB) has been set at 24.7 with a $\lambda_{\mathrm{eff}} = 4830$ \r{A} following \citet{chase_kilonova_2021}. Labels of the event are in the format (\textit{NS mass$\times$10})ns(\textit{BH mass})bh(\textit{effective spin$\times$10})$\,$c\textunderscore$X$s where if $X$ is 1 we use the standard deviations inferred from LV analysis, if it is 05 they are halved.}
\label{tab:confidence}
\end {table}

In order to take into account the predictable increase in the future sensitivity of LV detectors, we have considered in our analyses the possibility that the current average error (at the origin of the correlation between estimated masses and spin) will be halved.
As shown in the table, the augmented precision makes it slightly more easy to discriminate among the various EoSs, but the improvement is not very significant.

The most important result of our analysis is that, if the BH spin is not always vanishing, it is possible to discriminate among the various EoSs and, even more clearly, between the one-family and the two-families scenario. 
As shown in the Table (and also taking into account the dependence of $M_\mathrm{dyn}$ on the mass of the NS, shown in figure \ref{fig:mdyn_vs_mns}), a rather strong KN signal is expected in the one-family scenario, in particular for $M_\mathrm {NS}\sim (1.2-1.3) \mathrm{\,M}_\odot$ and $M_\mathrm{BH}\lesssim 5 \mathrm{\,M}_\odot$, if $\chi_\mathrm{BH}\gtrsim 0.2$. Moreover, if $M_\mathrm{BH}\lesssim 4 \mathrm{\,M}_\odot$ (as in the case of GW190425 \citep{gw190425}) a strong KN signal is expected in the one-family scenario even for a non-rotating BH. Instead, in the two-families scenario and in most of the analyzed cases almost no mass escapes the BH. Notice also that the hadronic EoS we have used in our analysis is not the softest possible. In previous papers (see e.g. \citet{drago_can_2014}) we have discussed even softer EoSs which can explain HSs with even smaller radii, such as those suggested in \citet{Ozel:2016oaf}, if their existence is confirmed.
Therefore, an even weaker KN signal can be justified within the two-families scenario.

A caveat is in order: in our analysis we have assumed that no matter is ejected in a QS-BH merger, as suggested by \citet{Kluzniak:2002dm}.
However, in that simulation the gravity of the BH is modelled through a pseudo-Newtonian potential with an absorbing boundary corresponding to the radius of the photon orbit in Schwarzschild geometry. The results of \citet{Kluzniak:2002dm} therefore need to be confirmed by new and more sophisticated numerical simulations in full GR. On the other hand, our main prediction is that for compact stars having masses about $(1.2-1.3) \mathrm{\,M}_\odot$, which in the two families scenario are HSs,
the KN signal is significantly suppressed respect to the one-family case and this prediction does not depend on the QS-BH simulation.

Obviously, also in NS-NS mergers it is possible to find clear signatures of the two-families scenario. As discussed in \cite{drago_merger_2018,pietri_merger_2019}, in the two-families scenario there are three types of mergers, namely HS-HS, HS-QS and QS-QS, and therefore the related phenomenology is very rich. For instance, the threshold mass for obtaining a prompt collapse to a BH depends on the type of merger. While in the one-family scenario one does not expect to have a prompt collapse for masses smaller than the mass of the binary at the origin of GW170817 ($\sim ~ 2.74$ M$_{\odot}$), in the two families scenario (in which GW170817 was associated with a HS-QS merger) there could be a prompt collapse for masses just above $2.5\mathrm{\,M}_\odot$, if the binary is made of two HSs, see \cite{pietri_merger_2019}. This is a unique prediction of the two-families scenario.

In conclusion, we have shown that the observation of the KN signal produced in a NS-BH merger can provide clear indications in favor or against the two-families scenario. The strongest discrimination between the "normal" scenario and the two-families scenario comes from mergers in which the masses of both the NS and the BH are rather small and $\chi_\mathrm{BH}\gtrsim 0.2$, and these constraints reduce the number of mergers which can be used in the analysis. However, the
next generation of telescopes and the next Ligo-VIRGO runs will be able to observe KN signals up to $\sim$~475~Mpc \citep{chase_kilonova_2021}, thus significantly enlarging the observable volume. The analysis just released by \citet{LIGOScientific:2021qlt} of two NS-BH mergers, GW200105 and GW200115, found that a small value of ejected mass was expected from both events, consistent with the absence of any EM counterpart. That same conclusion can be reached by analyzing the results we obtained in Figs. \ref{fig:mpa1sfho} and \ref{fig:504} and using the estimated masses and spins of those two events. On the other hand, as suggested in \citet{LIGOScientific:2021qlt}, the detection of those two mergers indicates that the estimated rate for this type of event is realistic and therefore we can expect, in the near future, to be able to gather crucial information from the search of KN signals from NS-BH mergers, unless the spin of all the BHs in a NS-BH merger is close to zero.

\bibliographystyle{aasjournal}

\begin{thebibliography}{}
\expandafter\ifx\csname natexlab\endcsname\relax\def\natexlab#1{#1}\fi
\providecommand{\url}[1]{\href{#1}{#1}}
\providecommand{\dodoi}[1]{doi:~\href{http://doi.org/#1}{\nolinkurl{#1}}}
\providecommand{\doeprint}[1]{\href{http://ascl.net/#1}{\nolinkurl{http://ascl.net/#1}}}
\providecommand{\doarXiv}[1]{\href{https://arxiv.org/abs/#1}{\nolinkurl{https://arxiv.org/abs/#1}}}

\bibitem[{Abbott {et~al.}(2018)}]{Abbott:2018exr}
{Abbott}, B.~P., {Abbott}, R., {Abbott}, T.~D., {et~al.} 2018, Phys. Rev. Lett., 121, 161101,
  \dodoi{10.1103/PhysRevLett.121.161101}

\bibitem[{{Abbott} {et~al.}(2020){Abbott}, {Abbott}, {Abbott}, {Abraham},
  {Acernese}, {Ackley}, {Adams}, {Adhikari}, {Adya}, {Affeldt}, {Agathos},
  {Agatsuma}, {Aggarwal}, {Aguiar}, {Aiello}, {Ain}, {Ajith}, {Allen},
  {Allocca}, {Aloy}, {Altin}, {Amato}, {Anand}, {Ananyeva}, {Anderson},
  {Anderson}, {Angelova}, {Antier}, {Appert}, {Arai}, {Araya}, {Areeda},
  {Ar{\`e}ne}, {Arnaud}, {Aronson}, {Arun}, {Ascenzi}, {Ashton}, {Aston},
  {Astone}, {Aubin}, {Aufmuth}, {AultONeal}, {Austin}, {Avendano},
  {Avila-Alvarez}, {Babak}, {Bacon}, {Badaracco}, {Bader}, {Bae}, {Baird},
  {Baker}, {Baldaccini}, {Ballardin}, {Ballmer}, {Bals}, {Banagiri},
  {Barayoga}, {Barbieri}, {Barclay}, {Barish}, {Barker}, {Barkett}, {Barnum},
  {Barone}, {Barr}, {Barsotti}, {Barsuglia}, {Barta}, {Bartlett}, {Bartos},
  {Bassiri}, {Basti}, {Bawaj}, {Bayley}, {Baylor}, {Bazzan}, {B{\'e}csy},
  {Bejger}, {Belahcene}, {Bell}, {Beniwal}, {Benjamin}, {Berger}, {Bergmann},
  {Bernuzzi}, {Berry}, {Bersanetti}, {Bertolini}, {Betzwieser}, {Bhandare},
  {Bidler}, {Biggs}, {Bilenko}, {Bilgili}, {Billingsley}, {Birney},
  {Birnholtz}, {Biscans}, {Bischi}, {Biscoveanu}, {Bisht}, {Bitossi},
  {Bizouard}, {Blackburn}, {Blackman}, {Blair}, {Blair}, {Blair}, {Bloemen},
  {Bobba}, {Bode}, {Boer}, {Boetzel}, {Bogaert}, {Bondu}, {Bonnand}, {Booker},
  {Boom}, {Bork}, {Boschi}, {Bose}, {Bossilkov}, {Bosveld}, {Bouffanais},
  {Bozzi}, {Bradaschia}, {Brady}, {Bramley}, {Branchesi}, {Brau}, {Breschi},
  {Briant}, {Briggs}, {Brighenti}, {Brillet}, {Brinkmann}, {Brockill},
  {Brooks}, {Brooks}, {Brown}, {Brunett}, {Buikema}, {Bulik}, {Bulten},
  {Buonanno}, {Buskulic}, {Buy}, {Byer}, {Cabero}, {Cadonati}, {Cagnoli},
  {Cahillane}, {Calder{\'o}n Bustillo}, {Callister}, {Calloni}, {Camp},
  {Campbell}, {Canepa}, {Cannon}, {Cao}, {Cao}, {Carapella}, {Carbognani},
  {Caride}, {Carney}, {Carullo}, {Casanueva Diaz}, {Casentini}, {Caudill},
  {Cavagli{\`a}}, {Cavalier}, {Cavalieri}, {Cella}, {Cerd{\'a}-Dur{\'a}n},
  {Cesarini}, {Chaibi}, {Chakravarti}, {Chamberlin}, {Chan}, {Chao},
  {Charlton}, {Chase}, {Chassande-Mottin}, {Chatterjee}, {Chaturvedi},
  {Chatziioannou}, {Cheeseboro}, {Chen}, {Chen}, {Chen}, {Cheng}, {Cheong},
  {Chia}, {Chiadini}, {Chincarini}, {Chiummo}, {Cho}, {Cho}, {Cho},
  {Christensen}, {Chu}, {Chua}, {Chung}, {Chung}, {Ciani}, {Cie{\'s}lar},
  {Ciobanu}, {Ciolfi}, {Cipriano}, {Cirone}, {Clara}, {Clark}, {Clearwater},
  {Cleva}, {Coccia}, {Cohadon}, {Cohen}, {Colleoni}, {Collette}, {Collins},
  {Colpi}, {Cominsky}, {Constancio}, {Conti}, {Cooper}, {Corban}, {Corbitt},
  {Cordero-Carri{\'o}n}, {Corezzi}, {Corley}, {Cornish}, {Corre}, {Corsi},
  {Cortese}, {Costa}, {Cotesta}, {Coughlin}, {Coughlin}, {Coulon},
  {Countryman}, {Couvares}, {Covas}, {Cowan}, {Coward}, {Cowart}, {Coyne},
  {Coyne}, {Creighton}, {Creighton}, {Cripe}, {Croquette}, {Crowder}, {Cullen},
  {Cumming}, {Cunningham}, {Cuoco}, {Dal Canton}, {D{\'a}lya}, {D'Angelo},
  {Danilishin}, {D'Antonio}, {Danzmann}, {Dasgupta}, {Da Silva Costa},
  {Datrier}, {Dattilo}, {Dave}, {Davier}, {Davis}, {Daw}, {DeBra},
  {Deenadayalan}, {Degallaix}, {De Laurentis}, {Del{\'e}glise}, {De Lillo},
  {Del Pozzo}, {DeMarchi}, {Demos}, {Dent}, {De Pietri}, {De Rosa}, {De Rossi},
  {DeSalvo}, {de Varona}, {Dhurandhar}, {D{\'\i}az}, {Dietrich}, {Di Fiore},
  {DiFronzo}, {Di Giorgio}, {Di Giovanni}, {Di Giovanni}, {Di Girolamo}, {Di
  Lieto}, {Ding}, {Di Pace}, {Di Palma}, {Di Renzo}, {Divakarla}, {Dmitriev},
  {Doctor}, {Donovan}, {Dooley}, {Doravari}, {Dorrington}, {Downes}, {Drago},
  {Driggers}, {Du}, {Ducoin}, {Dudi}, {Dupej}, {Durante}, {Dwyer}, {Easter},
  {Eddolls}, {Edo}, {Effler}, {Ehrens}, {Eichholz}, {Eikenberry}, {Eisenmann},
  {Eisenstein}, {Errico}, {Essick}, {Estelles}, {Estevez}, {Etienne}, {Etzel},
  {Evans}, {Evans}, {Fafone}, {Fairhurst}, {Fan}, {Farinon}, {Farr}, {Farr},
  {Fauchon-Jones}, {Favata}, {Fays}, {Fazio}, {Fee}, {Feicht}, {Fejer}, {Feng},
  {Fernandez-Galiana}, {Ferrante}, {Ferreira}, {Ferreira}, {Fidecaro}, {Fiori},
  {Fiorucci}, {Fishbach}, {Fisher}, {Fishner}, {Fittipaldi}, {Fitz-Axen},
  {Fiumara}, {Flaminio}, {Fletcher}, {Floden}, {Flynn}, {Fong}, {Font},
  {Forsyth}, {Fournier}, {Vivanco}, {Frasca}, {Frasconi}, {Frei}, {Freise},
  {Frey}, {Frey}, {Fritschel}, {Frolov}, {Fronz{\`e}}, {Fulda}, {Fyffe},
  {Gabbard}, {Gadre}, {Gaebel}, {Gair}, {Gamba}, {Gammaitoni}, {Gaonkar},
  {Garc{\'\i}a-Quir{\'o}s}, {Garufi}, {Gateley}, {Gaudio}, {Gaur}, {Gayathri},
  {Gemme}, {Genin}, {Gennai}, {George}, {George}, {George}, {Gergely},
  {Ghonge}, {Ghosh}, {Ghosh}, {Ghosh}, {Giacomazzo}, {Giaime}, {Giardina},
  {Gibson}, {Gill}, {Glover}, {Gniesmer}, {Godwin}, {Goetz}, {Goetz},
  {Goncharov}, {Gonz{\'a}lez}, {Castro}, {Gopakumar}, {Gossan}, {Gosselin},
  {Gouaty}, {Grace}, {Grado}, {Granata}, {Grant}, {Gras}, {Grassia}, {Gray},
  {Gray}, {Greco}, {Green}, {Green}, {Gretarsson}, {Grimaldi}, {Grimm},
  {Groot}, {Grote}, {Grunewald}, {Gruning}, {Guidi}, {Gulati}, {Guo}, {Gupta},
  {Gupta}, {Gupta}, {Gustafson}, {Gustafson}, {Haegel}, {Halim}, {Hall},
  {Hall}, {Hamilton}, {Hammond}, {Haney}, {Hanke}, {Hanks}, {Hanna}, {Hannam},
  {Hannuksela}, {Hansen}, {Hanson}, {Harder}, {Hardwick}, {Haris}, {Harms},
  {Harry}, {Harry}, {Hasskew}, {Haster}, {Haughian}, {Hayes}, {Healy},
  {Heidmann}, {Heintze}, {Heitmann}, {Hellman}, {Hello}, {Hemming}, {Hendry},
  {Heng}, {Hennig}, {Heurs}, {Hild}, {Hinderer}, {Ho}, {Hochheim}, {Hofman},
  {Holgado}, {Holland}, {Holt}, {Holz}, {Hopkins}, {Horst}, {Hough}, {Howell},
  {Hoy}, {Huang}, {H{\"u}bner}, {Huerta}, {Huet}, {Hughey}, {Hui}, {Husa},
  {Huttner}, {Huynh-Dinh}, {Idzkowski}, {Iess}, {Inchauspe}, {Ingram}, {Inta},
  {Intini}, {Irwin}, {Isa}, {Isac}, {Isi}, {Iyer}, {Jacqmin}, {Jadhav}, {Jani},
  {Janthalur}, {Jaranowski}, {Jariwala}, {Jenkins}, {Jiang}, {Johnson},
  {Johnson-McDaniel}, {Jones}, {Jones}, {Jones}, {Jones}, {Jonker}, {Ju},
  {Junker}, {Kalaghatgi}, {Kalogera}, {Kamai}, {Kandhasamy}, {Kang}, {Kanner},
  {Kapadia}, {Karki}, {Kashyap}, {Kasprzack}, {Kastaun}, {Katsanevas},
  {Katsavounidis}, {Katzman}, {Kaufer}, {Kawabe}, {Keerthana},
  {K{\'e}f{\'e}lian}, {Keitel}, {Kennedy}, {Key}, {Khalili}, {Khan}, {Khan},
  {Khazanov}, {Khetan}, {Khursheed}, {Kijbunchoo}, {Kim}, {Kim}, {Kim}, {Kim},
  {Kim}, {Kim}, {Kimball}, {King}, {Kinley-Hanlon}, {Kirchhoff}, {Kissel},
  {Kleybolte}, {Klika}, {Klimenko}, {Knowles}, {Koch}, {Koehlenbeck},
  {Koekoek}, {Koley}, {Kondrashov}, {Kontos}, {Koper}, {Korobko}, {Korth},
  {Kovalam}, {Kozak}, {Kr{\"a}mer}, {Kringel}, {Krishnendu}, {Kr{\'o}lak},
  {Krupinski}, {Kuehn}, {Kumar}, {Kumar}, {Kumar}, {Kumar}, {Kuo}, {Kutynia},
  {Kwang}, {Lackey}, {Laghi}, {Lai}, {Lam}, {Landry}, {Landry}, {Lane}, {Lang},
  {Lange}, {Lantz}, {Lanza}, {Lartaux-Vollard}, {Lasky}, {Laxen}, {Lazzarini},
  {Lazzaro}, {Leaci}, {Leavey}, {Lecoeuche}, {Lee}, {Lee}, {Lee}, {Lee}, {Lee},
  {Lee}, {Lehmann}, {Lenon}, {Leroy}, {Letendre}, {Levin}, {Li}, {Li}, {Li},
  {Li}, {Li}, {Lin}, {Linde}, {Linker}, {Littenberg}, {Liu}, {Liu},
  {Llorens-Monteagudo}, {Lo}, {London}, {Longo}, {Lorenzini}, {Loriette},
  {Lormand}, {Losurdo}, {Lough}, {Lousto}, {Lovelace}, {Lower}, {Lucaccioni},
  {L{\"u}ck}, {Lumaca}, {Lundgren}, {Lynch}, {Ma}, {Macas}, {Macfoy},
  {MacInnis}, {Macleod}, {Macquet}, {Maga{\~n}a Hernandez},
  {Maga{\~n}a-Sandoval}, {Magee}, {Majorana}, {Maksimovic}, {Malik}, {Man},
  {Mandic}, {Mangano}, {Mansell}, {Manske}, {Mantovani}, {Mapelli},
  {Marchesoni}, {Marion}, {M{\'a}rka}, {M{\'a}rka}, {Markakis}, {Markosyan},
  {Markowitz}, {Maros}, {Marquina}, {Marsat}, {Martelli}, {Martin}, {Martin},
  {Martinez}, {Martynov}, {Masalehdan}, {Mason}, {Massera}, {Masserot},
  {Massinger}, {Masso-Reid}, {Mastrogiovanni}, {Matas}, {Matichard}, {Matone},
  {Mavalvala}, {McCann}, {McCarthy}, {McClelland}, {McCormick}, {McCuller},
  {McGuire}, {McIsaac}, {McIver}, {McManus}, {McRae}, {McWilliams}, {Meacher},
  {Meadors}, {Mehmet}, {Mehta}, {Meidam}, {Mejuto Villa}, {Melatos}, {Mendell},
  {Mercer}, {Mereni}, {Merfeld}, {Merilh}, {Merzougui}, {Meshkov}, {Messenger},
  {Messick}, {Messina}, {Metzdorff}, {Meyers}, {Meylahn}, {Miani}, {Miao},
  {Michel}, {Middleton}, {Milano}, {Miller}, {Millhouse}, {Mills},
  {Milovich-Goff}, {Minazzoli}, {Minenkov}, {Mishkin}, {Mishra}, {Mistry},
  {Mitra}, {Mitrofanov}, {Mitselmakher}, {Mittleman}, {Mo}, {Moffa}, {Mogushi},
  {Mohapatra}, {Molina-Ruiz}, {Mondin}, {Montani}, {Moore}, {Moraru},
  {Morawski}, {Moreno}, {Morisaki}, {Mours}, {Mow-Lowry}, {Muciaccia},
  {Mukherjee}, {Mukherjee}, {Mukherjee}, {Mukherjee}, {Mukund}, {Mullavey},
  {Munch}, {Mu{\~n}iz}, {Muratore}, {Murray}, {Nagar}, {Nardecchia},
  {Naticchioni}, {Nayak}, {Neil}, {Neilson}, {Nelemans}, {Nelson}, {Nery},
  {Neunzert}, {Nevin}, {Ng}, {Ng}, {Nguyen}, {Nguyen}, {Nichols}, {Nichols},
  {Nissanke}, {Nocera}, {North}, {Nuttall}, {Obergaulinger}, {Oberling},
  {O'Brien}, {Oganesyan}, {Ogin}, {Oh}, {Oh}, {Ohme}, {Ohta}, {Okada},
  {Oliver}, {Oppermann}, {Oram}, {O'Reilly}, {Ormiston}, {Ortega},
  {O'Shaughnessy}, {Ossokine}, {Ottaway}, {Overmier}, {Owen}, {Pace}, {Pagano},
  {Page}, {Pagliaroli}, {Pai}, {Pai}, {Palamos}, {Palashov}, {Palomba}, {Pan},
  {Panda}, {Pang}, {Pankow}, {Pannarale}, {Pant}, {Paoletti}, {Paoli},
  {Parida}, {Parker}, {Pascucci}, {Pasqualetti}, {Passaquieti}, {Passuello},
  {Patil}, {Patricelli}, {Payne}, {Pearlstone}, {Pechsiri}, {Pedersen},
  {Pedraza}, {Pedurand}, {Pele}, {Penn}, {Perego}, {Perez}, {P{\'e}rigois},
  {Perreca}, {Petermann}, {Pfeiffer}, {Phelps}, {Phukon}, {Piccinni}, {Pichot},
  {Piergiovanni}, {Pierro}, {Pillant}, {Pinard}, {Pinto}, {Pirello}, {Pitkin},
  {Plastino}, {Poggiani}, {Pong}, {Ponrathnam}, {Popolizio}, {Porter},
  {Powell}, {Prajapati}, {Prasad}, {Prasai}, {Prasanna}, {Pratten},
  {Prestegard}, {Principe}, {Prodi}, {Prokhorov}, {Punturo}, {Puppo},
  {P{\"u}rrer}, {Qi}, {Quetschke}, {Quinonez}, {Raab}, {Raaijmakers},
  {Radkins}, {Radulesco}, {Raffai}, {Raja}, {Rajan}, {Rajbhandari},
  {Rakhmanov}, {Ramirez}, {Ramos-Buades}, {Rana}, {Rao}, {Rapagnani},
  {Raymond}, {Razzano}, {Read}, {Regimbau}, {Rei}, {Reid}, {Reitze},
  {Rettegno}, {Ricci}, {Richardson}, {Richardson}, {Ricker}, {Riemenschneider},
  {Riles}, {Rizzo}, {Robertson}, {Robinet}, {Rocchi}, {Rolland}, {Rollins},
  {Roma}, {Romanelli}, {Romano}, {Romel}, {Romie}, {Rose}, {Rose}, {Rose},
  {Rosell}, {Rosi{\'n}ska}, {Rosofsky}, {Ross}, {Rowan}, {Roy}, {R{\"u}diger},
  {Ruggi}, {Rutins}, {Ryan}, {Sachdev}, {Sadecki}, {Sakellariadou}, {Salafia},
  {Salconi}, {Saleem}, {Samajdar}, {Sammut}, {Sanchez}, {Sanchez},
  {Sanchis-Gual}, {Sanders}, {Santiago}, {Santos}, {Sarin}, {Sassolas},
  {Sathyaprakash}, {Sauter}, {Savage}, {Schale}, {Scheel}, {Scheuer},
  {Schmidt}, {Schnabel}, {Schofield}, {Sch{\"o}nbeck}, {Schreiber}, {Schulte},
  {Schutz}, {Scott}, {Scott}, {Seidel}, {Sellers}, {Sengupta}, {Sennett},
  {Sentenac}, {Sequino}, {Sergeev}, {Setyawati}, {Shaddock}, {Shaffer},
  {Shahriar}, {Shaner}, {Sharma}, {Sharma}, {Shawhan}, {Shen}, {Shink},
  {Shoemaker}, {Shoemaker}, {Shukla}, {ShyamSundar}, {Siellez}, {Sieniawska},
  {Sigg}, {Singer}, {Singh}, {Singh}, {Singhal}, {Sintes}, {Sitmukhambetov},
  {Skliris}, {Slagmolen}, {Slaven-Blair}, {Smith}, {Smith}, {Somala}, {Son},
  {Soni}, {Sorazu}, {Sorrentino}, {Souradeep}, {Sowell}, {Spencer}, {Spera},
  {Srivastava}, {Srivastava}, {Staats}, {Stachie}, {Standke}, {Steer},
  {Steinke}, {Steinlechner}, {Steinlechner}, {Steinmeyer}, {Stevenson},
  {Stocks}, {Stone}, {Stops}, {Strain}, {Stratta}, {Strigin}, {Strunk},
  {Sturani}, {Stuver}, {Sudhir}, {Summerscales}, {Sun}, {Sunil}, {Sur},
  {Suresh}, {Sutton}, {Swinkels}, {Szczepa{\'n}czyk}, {Tacca}, {Tait},
  {Talbot}, {Tanner}, {Tao}, {T{\'a}pai}, {Tapia}, {Tasson}, {Taylor},
  {Tenorio}, {Terkowski}, {Thomas}, {Thomas}, {Thondapu}, {Thorne}, {Thrane},
  {Tiwari}, {Tiwari}, {Tiwari}, {Toland}, {Tonelli}, {Tornasi},
  {Torres-Forn{\'e}}, {Torrie}, {T{\"o}yr{\"a}}, {Travasso}, {Traylor},
  {Tringali}, {Tripathee}, {Trovato}, {Trozzo}, {Tsang}, {Tse}, {Tso},
  {Tsukada}, {Tsuna}, {Tsutsui}, {Tuyenbayev}, {Ueno}, {Ugolini},
  {Unnikrishnan}, {Urban}, {Usman}, {Vahlbruch}, {Vajente}, {Valdes},
  {Valentini}, {van Bakel}, {van Beuzekom}, {van den Brand}, {Van Den Broeck},
  {Vander-Hyde}, {van der Schaaf}, {VanHeijningen}, {van Veggel}, {Vardaro},
  {Varma}, {Vass}, {Vas{\'u}th}, {Vecchio}, {Vedovato}, {Veitch}, {Veitch},
  {Venkateswara}, {Venugopalan}, {Verkindt}, {Vetrano}, {Vicer{\'e}}, {Viets},
  {Vinciguerra}, {Vine}, {Vinet}, {Vitale}, {Vo}, {Vocca}, {Vorvick},
  {Vyatchanin}, {Wade}, {Wade}, {Wade}, {Walet}, {Walker}, {Wallace}, {Walsh},
  {Wang}, {Wang}, {Wang}, {Wang}, {Ward}, {Warden}, {Warner}, {Was}, {Watchi},
  {Weaver}, {Wei}, {Weinert}, {Weinstein}, {Weiss}, {Wellmann}, {Wen},
  {Wessel}, {We{\ss}els}, {Westhouse}, {Wette}, {Whelan}, {White}, {Whiting},
  {Whittle}, {Wilken}, {Williams}, {Williamson}, {Willis}, {Willke}, {Winkler},
  {Wipf}, {Wittel}, {Woan}, {Woehler}, {Wofford}, {Wright}, {Wu}, {Wysocki},
  {Xiao}, {Xu}, {Yamamoto}, {Yancey}, {Yang}, {Yang}, {Yang}, {Yap}, {Yazback},
  {Yeeles}, {Yu}, {Yu}, {Yuen}, {Zadro{\.z}ny}, {Zadro{\.z}ny}, {Zanolin},
  {Zelenova}, {Zendri}, {Zevin}, {Zhang}, {Zhang}, {Zhang}, {Zhao}, {Zhao},
  {Zhou}, {Zhou}, {Zhu}, {Zimmerman}, {Zucker}, \& {Zweizig}}]{gw190425}
{Abbott}, B.~P., {Abbott}, R., {Abbott}, T.~D., {et~al.} 2020, \apjl, 892, L3,
  \dodoi{10.3847/2041-8213/ab75f5}

\bibitem[{Abbott {et~al.}(2021)Abbott, Abbott, Abraham, Acernese, Ackley,
  Adams, Adams, Adhikari, \& Adya}]{LIGOScientific:2021qlt}
Abbott, R., Abbott, T., Abraham, S., {et~al.} 2021, Astrophys. J. Lett., 915,
  L5, \dodoi{10.3847/2041-8213/ac082e}

\bibitem[{Akmal {et~al.}(1998)Akmal, Pandharipande, \& Ravenhall}]{AP3}
Akmal, A., Pandharipande, V.~R., \& Ravenhall, D.~G. 1998, Phys. Rev. C, 58,
  1804, \dodoi{10.1103/PhysRevC.58.1804}

\bibitem[{Barbieri {et~al.}(2020)Barbieri, Salafia, Perego, Colpi, \&
  Ghirlanda}]{Barbieri2020}
Barbieri, C., Salafia, O.~S., Perego, A., Colpi, M., \& Ghirlanda, G. 2020, The
  European Physical Journal A, 56, 8, \dodoi{10.1140/epja/s10050-019-00013-x}

\bibitem[{{Bardeen} {et~al.}(1972){Bardeen}, {Press}, \&
  {Teukolsky}}]{bardeen1972}
{Bardeen}, J.~M., {Press}, W.~H., \& {Teukolsky}, S.~A. 1972, \apj, 178, 347,
  \dodoi{10.1086/151796}

\bibitem[{{Belczynski} {et~al.}(2008){Belczynski}, {Kalogera}, {Rasio}, {Taam},
  {Zezas}, {Bulik}, {Maccarone}, \& {Ivanova}}]{Belczynski_2008}
{Belczynski}, K., {Kalogera}, V., {Rasio}, F.~A., {et~al.} 2008, \apjs, 174,
  223, \dodoi{10.1086/521026}

\bibitem[{{Belczynski} {et~al.}(2020){Belczynski}, {Klencki}, {Fields},
  {Olejak}, {Berti}, {Meynet}, {Fryer}, {Holz}, {O'Shaughnessy}, {Brown},
  {Bulik}, {Leung}, {Nomoto}, {Madau}, {Hirschi}, {Kaiser}, {Jones}, {Mondal},
  {Chruslinska}, {Drozda}, {Gerosa}, {Doctor}, {Giersz}, {Ekstrom}, {Georgy},
  {Askar}, {Baibhav}, {Wysocki}, {Natan}, {Farr}, {Wiktorowicz}, {Coleman
  Miller}, {Farr}, \& {Lasota}}]{Belczynski_2020}
{Belczynski}, K., {Klencki}, J., {Fields}, C.~E., {et~al.} 2020, \aap, 636,
  A104, \dodoi{10.1051/0004-6361/201936528}

\bibitem[{Bombaci {et~al.}(2021)Bombaci, Drago, Logoteta, Pagliara, \&
  Vidaña}]{bombaci_was_2021}
Bombaci, I., Drago, A., Logoteta, D., Pagliara, G., \& Vidaña, I. 2021, Phys.
  Rev. Lett., 126, 162702, \dodoi{10.1103/PhysRevLett.126.162702}

\bibitem[{Bombaci {et~al.}(2004)Bombaci, Parenti, \& Vidana}]{Bombaci_2004}
Bombaci, I., Parenti, I., \& Vidana, I. 2004, The Astrophysical Journal, 614,
  314, \dodoi{10.1086/423658}

\bibitem[{Burgio {et~al.}(2018)Burgio, Drago, Pagliara, Schulze, \&
  Wei}]{Burgio:2018yix}
Burgio, G.~F., Drago, A., Pagliara, G., Schulze, H.~J., \& Wei, J.~B. 2018,
  Astrophys. J., 860, 139, \dodoi{10.3847/1538-4357/aac6ee}

\bibitem[{Capano {et~al.}(2020)Capano, Tews, Brown, Margalit, De, Kumar, Brown,
  Krishnan, \& Reddy}]{Capano:2019eae}
Capano, C.~D., Tews, I., Brown, S.~M., {et~al.} 2020, Nature Astron., 4, 625,
  \dodoi{10.1038/s41550-020-1014-6}

\bibitem[{Chase {et~al.}(2022)Chase, O'Connor, Fryer, Troja, Korobkin,
  Wollaeger, Ristic, Fontes, Hungerford, \& Herring}]{chase_kilonova_2021}
Chase, E.~A., O'Connor, B., Fryer, C.~L., {et~al.} 2022, The Astrophysical
  Journal, 927, 163, \dodoi{10.3847/1538-4357/ac3d25}

\bibitem[{De~Pietri {et~al.}(2019)De~Pietri, Drago, Feo, Pagliara, Pasquali,
  Traversi, \& Wiktorowicz}]{pietri_merger_2019}
De~Pietri, R., Drago, A., Feo, A., {et~al.} 2019, ApJ, 881, 122,
  \dodoi{10.3847/1538-4357/ab2fd0}

\bibitem[{Drago {et~al.}(2014{\natexlab{a}})Drago, Lavagno, \&
  Pagliara}]{drago_can_2014}
Drago, A., Lavagno, A., \& Pagliara, G. 2014{\natexlab{a}}, Phys. Rev. D, 89,
  043014, \dodoi{10.1103/PhysRevD.89.043014}

\bibitem[{Drago {et~al.}(2014{\natexlab{b}})Drago, Lavagno, Pagliara, \&
  Pigato}]{PhysRevC.90.065809}
Drago, A., Lavagno, A., Pagliara, G., \& Pigato, D. 2014{\natexlab{b}}, Phys.
  Rev. C, 90, 065809, \dodoi{10.1103/PhysRevC.90.065809}

\bibitem[{Drago \& Pagliara(2018)}]{drago_merger_2018}
Drago, A., \& Pagliara, G. 2018, ApJ, 852, L32,
  \dodoi{10.3847/2041-8213/aaa40a}

\bibitem[{Drago \& Pagliara(2020)}]{Drago:2020gqn}
---. 2020, Phys. Rev. D, 102, 063003, \dodoi{10.1103/PhysRevD.102.063003}

\bibitem[{Foucart {et~al.}(2018)Foucart, Hinderer, \& Nissanke}]{Foucart2018}
Foucart, F., Hinderer, T., \& Nissanke, S. 2018, Phys. Rev. D, 98, 081501,
  \dodoi{10.1103/PhysRevD.98.081501}

\bibitem[{Fragione(2021)}]{Fragione_2021}
Fragione, G. 2021, The Astrophysical Journal Letters, 923, L2,
  \dodoi{10.3847/2041-8213/ac3bcd}

\bibitem[{Kawaguchi {et~al.}(2016)Kawaguchi, Kyutoku, Shibata, \&
  Tanaka}]{Kawaguchi2016}
Kawaguchi, K., Kyutoku, K., Shibata, M., \& Tanaka, M. 2016, The Astrophysical
  Journal, 825, 52, \dodoi{10.3847/0004-637x/825/1/52}

\bibitem[{Kluzniak \& Lee(2002)}]{Kluzniak:2002dm}
Kluzniak, W., \& Lee, W.~H. 2002, Mon. Not. Roy. Astron. Soc., 335, L29,
  \dodoi{10.1046/j.1365-8711.2002.05819.x}

\bibitem[{Liu \& Lai(2021)}]{10.1093/mnras/stab178}
Liu, B., \& Lai, D. 2021, Monthly Notices of the Royal Astronomical Society,
  502, 2049, \dodoi{10.1093/mnras/stab178}

\bibitem[{Lu {et~al.}(2020)Lu, Beniamini, \& Bonnerot}]{10.1093/mnras/staa3372}
Lu, W., Beniamini, P., \& Bonnerot, C. 2020, Monthly Notices of the Royal
  Astronomical Society, 500, 1817, \dodoi{10.1093/mnras/staa3372}

\bibitem[{Markakis {et~al.}(2009)Markakis, Read, Shibata, Ury{\={u}},
  Creighton, Friedman, \& Lackey}]{Markakis_2009}
Markakis, C., Read, J.~S., Shibata, M., {et~al.} 2009, Journal of Physics:
  Conference Series, 189, 012024, \dodoi{10.1088/1742-6596/189/1/012024}

\bibitem[{Miller {et~al.}(2021)Miller, Lamb, Dittmann, Bogdanov, Arzoumanian,
  Gendreau, Guillot, Ho, Lattimer, Loewenstein, Morsink, Ray, Wolff, Baker,
  Cazeau, Manthripragada, Markwardt, Okajima, Pollard, Cognard, Cromartie,
  Fonseca, Guillemot, Kerr, Parthasarathy, Pennucci, Ransom, \&
  Stairs}]{miller2021nicer}
Miller, M.~C., Lamb, F.~K., Dittmann, A.~J., {et~al.} 2021, The Astrophysical
  Journal Letters, 918, L28, \dodoi{10.3847/2041-8213/ac089b}

\bibitem[{Most {et~al.}(2018)Most, Weih, Rezzolla, \&
  Schaffner-Bielich}]{Most:2018hfd}
Most, E.~R., Weih, L.~R., Rezzolla, L., \& Schaffner-Bielich, J. 2018, Phys.
  Rev. Lett., 120, 261103, \dodoi{10.1103/PhysRevLett.120.261103}

\bibitem[{Müther {et~al.}(1987)Müther, Prakash, \& Ainsworth}]{MUTHER1987469}
Müther, H., Prakash, M., \& Ainsworth, T. 1987, Physics Letters B, 199, 469,
  \dodoi{https://doi.org/10.1016/0370-2693(87)91611-X}

\bibitem[{N\"attil\"a {et~al.}(2017)N\"attil\"a, Miller, Steiner, Kajava,
  Suleimanov, \& Poutanen}]{Nattila:2017wtj}
N\"attil\"a, J., Miller, M.~C., Steiner, A.~W., {et~al.} 2017, Astron.
  Astrophys., 608, A31, \dodoi{10.1051/0004-6361/201731082}

\bibitem[{{Ng} {et~al.}(2018){Ng}, {Vitale}, {Zimmerman}, {Chatziioannou},
  {Gerosa}, \& {Haster}}]{ngetal}
{Ng}, K. K.~Y., {Vitale}, S., {Zimmerman}, A., {et~al.} 2018, \prd, 98, 083007,
  \dodoi{10.1103/PhysRevD.98.083007}

\bibitem[{\"Ozel \& Freire(2016)}]{Ozel:2016oaf}
\"Ozel, F., \& Freire, P. 2016, Ann. Rev. Astron. Astrophys., 54, 401,
  \dodoi{10.1146/annurev-astro-081915-023322}

\bibitem[{Raaijmakers {et~al.}(2021)Raaijmakers, Greif, Hebeler, Hinderer,
  Nissanke, Schwenk, Riley, Watts, Lattimer, \& Ho}]{Raaijmakers:2021uju}
Raaijmakers, G., Greif, S.~K., Hebeler, K., {et~al.} 2021, The Astrophysical
  Journal Letters, 918, L29, \dodoi{10.3847/2041-8213/ac089a}

\bibitem[{Riley {et~al.}(2019)}]{Riley:2019yda}
Riley, T.~E., Watts, A.~L., Bogdanov, S., {et~al.} 2019, Astrophys. J. Lett., 887, L21,
  \dodoi{10.3847/2041-8213/ab481c}

\bibitem[{Riley {et~al.}(2021)Riley, Watts, Ray, Bogdanov, Guillot, Morsink,
  Bilous, Arzoumanian, Choudhury, Deneva, Gendreau, Harding, Ho, Lattimer,
  Loewenstein, Ludlam, Markwardt, Okajima, Prescod-Weinstein, Remillard, Wolff,
  Fonseca, Cromartie, Kerr, Pennucci, Parthasarathy, Ransom, Stairs, Guillemot,
  \& Cognard}]{riley2021nicer}
Riley, T.~E., Watts, A.~L., Ray, P.~S., {et~al.} 2021, The Astrophysical
  Journal Letters, 918, L27, \dodoi{10.3847/2041-8213/ac0a81}

\bibitem[{Shibata \& Taniguchi(2008)}]{Shibata:2007zm}
Shibata, M., \& Taniguchi, K. 2008, Phys. Rev. D, 77, 084015,
  \dodoi{10.1103/PhysRevD.77.084015}

\bibitem[{{Steiner} {et~al.}(2013){Steiner}, {Hempel}, \& {Fischer}}]{sfho}
{Steiner}, A.~W., {Hempel}, M., \& {Fischer}, T. 2013, \apj, 774, 17,
  \dodoi{10.1088/0004-637X/774/1/17}

\bibitem[{Traversi {et~al.}(2021)Traversi, Char, Pagliara, \&
  Drago}]{Traversi:2021fad}
Traversi, S., Char, P., Pagliara, G., \& Drago, A. 2021.
\newblock \doarXiv{2102.02357}

\bibitem[{Typel {et~al.}(2010)Typel, R\"opke, Kl\"ahn, Blaschke, \&
  Wolter}]{typelDD2}
Typel, S., R\"opke, G., Kl\"ahn, T., Blaschke, D., \& Wolter, H.~H. 2010, Phys.
  Rev. C, 81, 015803, \dodoi{10.1103/PhysRevC.81.015803}

\bibitem[{Wiktorowicz(2021)}]{Wiktorowicz}
Wiktorowicz, G. 2021, private communication

\bibitem[{Zhu {et~al.}(2021)Zhu, Wu, Yang, Zhang, Yu, Gao, Cao, \&
  Liu}]{Zhu_2021}
Zhu, J.-P., Wu, S., Yang, Y.-P., {et~al.} 2021, The Astrophysical Journal, 921,
  156, \dodoi{10.3847/1538-4357/ac19a7}

\end{thebibliography}

\end{document}